\documentclass[11pt,a4paper]{article}
\usepackage{jheppub}

\newcount\xrefpos \xrefpos=0
\newcount\yrefpos \yrefpos=0
\newcount\xput \xput=0
\newcount\yput \yput=0

\newcommand{\be}{\begin{equation}}
\newcommand{\ee}{\end{equation}}

\newcommand{\bea}{\begin{eqnarray}}
\newcommand{\eea}{\end{eqnarray}}
\newcommand{\eref}[1]{Eq.~(\ref{#1})}
\newcommand{\nn}{\nonumber}

\def\2F1{\mbox{$_2${F}$_1$}}

\title{Low temperature properties of holographic condensates}
\author{Pallab Basu}
\affiliation{Department of Physics and Astronomy, \\
University of Kentucky, Lexington, KY 40506, USA}
\emailAdd{pallab.basu@uky.edu}

\abstract{In the current work we study various models of holographic superconductors at low temperature. Generically the zero temperature limit of those models are solitonic solution with a zero sized horizon. Here we generalized simple version of those zero temperature solutions to small but non-zero temperature $T$. We confine ourselves to cases where near horizon geometry is $AdS^4$. At a non-zero temperature a small horizon would form deep inside this $AdS^4$ which does not disturb the UV physics. The resulting geometry may be matched with the zero temperature solution at an intermediate length scale. We understand this matching from separation of scales by setting up a perturbative expansion in gauge potential. We have a better analytic control in abelian case and quantities may be expressed in terms of hypergeometric function. From this we calculate low temperature behavior of various quatities like entropy, charge density and specific heat etc. We also calculate various energy gaps associated with p-wave holographic superconductor to understand the underlying pairing mechanism. The result deviates significantly from the corresponding weak coupling BCS counterpart. }

\begin{document}

\maketitle

\section{Introduction}
In recent years string theory or, more specifically, gauge-gravity duality has seen interesting applications in the field of condensed matter physics. One of the earliest such applications is the discovery of a superconductor-like phase transition in AdS with a Reissner-Nordstr\"{o}m black hole and a charged scalar field minimally coupled to a $U(1)$ gauge field \cite{Gubser:2008px,Hartnoll:2008vx,Hartnoll:2008kx}. Other system with similar properties include  so-called ``p-wave'' holographic superconductors with non-abelian gauge fields instead of a scalar coupled to $U(1)$ gauge fields \cite{Gubser:2008wv,Gubser:2008zu,Roberts:2008ns}. In various works \cite{Horowitz:2009ij,Gubser:2009cg,Gubser:2009gp, Gubser:2008pf, Gauntlett:2009dn, Gubser:2008wz, Konoplya:2009hv,Basu:2009vv,Ammon:2009xh}, authors have studied related systems in the zero temperature limit. 


Generically the zero temperature solutions turn out to be a solitonic solution with a zero sized horizon. The authors find that the effective potential for small gauge field fluctuations vanishes near the black hole horizon for the abelian cases. This implies that the normal component of the A.C. conductivity never vanishes, even at zero temperature, which in turn indicates that the superconductor is gapless \cite{Horowitz:2009ij}. However the corresponding effective potential does not vanish at the horizon for the non-abelian case. It is concluded that the holographic non-abelian superconductor does have a finite gap for the relevant gauge field fluctuations \cite{Basu:2009vv}. The non-abelian system is an anisotropic system which shows different conductivity in different directions.

The near horizon geometry of those zero temperature solutions are interesting and ranges from simple $AdS^4$ to various complicated and Lifshitz like geometries. Near the horizon these geometries could be constructed by an analytic perturbation theory  \cite{Horowitz:2009ij,Basu:2009vv}. Near horizon values of the scalar field or the appropriate component of the gauge field enters as a undetermined parameter in those perturbative expansion. Those parameters are determined by a numerical integration to infinity and consequent application of proper boundary conditions. 

In this work we generalize simple version of those zero temperature solutions to small but non-zero temperature $T$. It should be noted that a non-zero temperature solution was already obtained numerically \cite{Hartnoll:2008kx,Horowitz:2009ij,Ammon:2009xh}. However from a purely numerical solution it is difficult to conclude about the low temperature analytic behaviour of various quantities. Whereas we will be able to calculate the nature of various interesting physical quantities analytically. We confine ourselves to cases where near horizon geometry is $AdS^4$. We expect that at non-zero temperature a small horizon would form deep inside this $AdS^4$. We intuitively understand this by separation of scales. As the black hole is situated deep inside the $AdS^4$ it does not affect the UV physics. Hence we expect that at an intermediate scale ($r_i$) the non-zero temperature solution approaches the zero temp one. Importantly for a very small horizon size ($r_0$) the intermediate scale may itself be chosen very small so that the zero temperature perturbative method would be valid in the intermediate scale. Here we have $r_0 \ll r_i \ll 1$. We show that we can  set up a perturbative expansion in terms of the gauge field $A_0$ which interpolates between the black hole horizon and the intermediate scale. From this matching in the intermediate scale we argue that a slight variation of the zero temperature numerics may be applied to the non-zero temperature case. 

From our solution we may calculate how entropy, specific heat etc. vanishes near zero temperature. We also calculate the various energy gap associated with the systems.  Especially in the non-abelian case we calculate the various energy gap in the system and from their ratio we find some hint of underlying ``pairing mechanism". The ratio deviates around $33\%$ from its weak coupling BCS counterpart.

Our results may be generalized to various cases where near horizon geometry at zero temperature deviates from $AdS^4$ \cite{Horowitz:2009ij,Gubser:2009cg,Gubser:2009gp, Gubser:2008pf, Gauntlett:2009dn, Gubser:2008wz}. The application part may include calculation of various fermionic propagator, calculation of second sound and more interestingly low temperature behaviour of non-universality of viscosity entropy ratio etc \cite{Herzog:2009ci,Chen:2009pt,Gubser:2009dt,Ammon:2010pg,Gubser:2010dm,Faulkner:2009am,Erdmenger:2010xm,Natsuume:2010ky}.

Plan of this paper is as follows. In section \ref{sec:ab} we discuss the abelian or s-wave case. In section \ref{sec:nonab} we discuss the non-abelian or p-wave case.
\section{Abelian holographic superconductors}
\label{sec:ab}
We begin with the following four dimensional action describing gravity minimally coupled to a Maxwell field and charged scalar:
\be\label{eq:bulktheory}
{\cal L} = R + \frac{6}{L^2} - \frac{1}{4} F^{\mu\nu} F_{\mu\nu} 
- |\nabla \psi - i q A \psi |^2 -V( |\psi|) \,.
\ee
As usual we are writing $F=dA$, the cosmological constant is $-3/L^2$, and $m,q$ are the mass and charge of the scalar field. We are interested in plane symmetric solutions, so we set
\be\label{metric}
 ds^2=-g(r) e^{-\chi(r)} dt^2+{dr^2\over g(r)}+r^2(dx^2+dy^2)
\ee
\be
A=A_0(r)~dt, \quad \psi = \psi(r)
\ee
We can choose a gauge in which $\psi$ is real and work in units with $L=1$. The equations of motion are:
\be \psi''+\left(\frac{g'}{g}-\frac{\chi'}{2}+\frac{2}{r} \right)\psi' +\frac{q^2A_0^2e^\chi}{g^2}  \psi  -{V'(\psi)\over 2g}=0\label{psieom}\ee

\be\label{phieom}
A_0''+\left(\frac{\chi'}{2}+\frac{2}{r}  \right)A_0'-\frac{2q^2\psi^2}{g}A_0=0
\ee

\be
\chi'+r\psi'^2+\frac{rq^2A_0^2\psi^2e^\chi}{g^2}=0\label{chieom}
\ee

\be\label{geom}
g' + \left(\frac{1}{r}  - { \chi'\over 2}\right) g+\frac{rA_0'^2e^\chi}{4}- 3r+\frac{rV(\psi)}{2}=0
\ee
These equations are invariant under a scaling symmetries:
\bea\label{rescales}
r \to a r \,, \quad (t,x,y) \to (t,x,y)/a \,, \quad g \to a^2 g \,, \quad A_0 \to a A_0 \\
e^\chi \to b^2 e^\chi, \quad  t\to bt, \quad A_0 \to A_0/b
\eea
Once a solution is found, this symmetry can be used to set $\chi =0$ at the boundary at infinity, so the metric takes the standard AdS form asymptotically.

At large radius
\be 
A_0 = \mu -{\rho\over r}, \qquad \psi ={\psi^{(\lambda)}\over r^\lambda}+{\psi^{(3-\lambda)}\over r^{3-\lambda}}.
\ee
where $\lambda = (3 +\sqrt{9+4m^2})/2 $. In the boundary CFT, $\mu$ is the chemical potential, $\rho$ is the charge density, and $\lambda$ is the scaling dimension of the operator dual to $\psi$. We want this operator to condense without being sourced, so we are only interested in solutions where $\psi$ is normalizable. This typically requires setting $\psi^{(3-\lambda)} = 0$. As the boundary chemical potential is increased beyond certain critical value, $\psi$ condenses. One may ask about the zero temperature limit of such a configuration. At $T=0$, a condensation of $\psi$ is possible only if $m^2-2 q^2 < -3/2$. 

\subsection{$m^2 = 0$}
 
Here we would rephrase the results of \cite{Horowitz:2009ij} in our terms. We like to find the superconducting ground state of the system with a non-zero condensate. We will confine ourselves to $m^2 = 0$ case. Being a single state without any degeneracy, a superconducting ground state does not have any entropy associated with it (\cite{Horowitz:2009ij},\cite{Gubser:2009cg,Gubser:2009gp}). 

 We start by guessing a near horizon ansatz, $g(r)=r^2,\psi = \psi_0$. We use the scaling symmetries \eref{rescales} to set the co-efficient in front of $g$ to unity. Once a suitable $g(r)$ is chosen, we set up a step by step perturbation in $A_0$. The first step is to solve for the eqn. of motion of $A_0$ in this metric,
\be
A_0 = r^{2+\alpha}, \quad q\psi_0 = \left({\alpha^2 + 5\alpha + 6\over 2}\right)^{1/2}
\ee
Here we have used the scaling symmetries to rescale the co-efficient of $A_0$ to $1$. All the other metric component and scalar field are kept $r$ independent at this step.

 In the next order in $A_0$ one may solve for the $r$ dependence of other fields (assuming $\alpha > -1$). This procedure works as long as the various $A_0$ dependent quantities appearing in the perturbative expansion are small.
We get,
\be
\quad \psi = \psi_0 - \psi_1(r), \quad \chi = - \chi_1(r), \quad
g=r^2 - g_1(r)
\ee
 where,
\be
\quad \chi_1 = {\alpha^2 + 5\alpha + 6\over 4(\alpha + 1)}e^{\chi_o}r^{2(1+\alpha)}
\ee
\be
g_1 =  {\alpha + 2\over 4} e^{\chi_o}r^{4+2\alpha}, \quad \psi_1 = {q e^{\chi_o} \over 2(2\alpha^2 + 7\alpha +5)}\left({\alpha^2 + 5\alpha + 6\over 2}\right)^{1/2} r^{2(1+\alpha)}.
\label{scaling}
\ee
This scaling solution is valid in the regime $r \ll 1$. This solution may be used as a boundary condition to the EOM's for a numerical integration to infinity. In general the value of $\psi(\infty)$ will be non-zero. The value of $\alpha$ is determined by the requirement $\psi(\infty)$ is zero. 

\subsection{Small non-zero $T$}

From the scaling relations \eref{scaling} one finds that IR geometry ($r \rightarrow 0$) is an emergent $AdS^4$ with the same cosmological constant as that of the boundary $AdS^4$. At small non-zero $T$ we guess that a black hole horizon will be created in the deep $IR$ region of the emergent $AdS^4$. Hence at the first step we choose the following ansatz,
\begin{align}
g(r)=r^2 \left(1-\frac{r_0^3}{r^3}\right), \quad \psi=\psi_0
\label{testmetric}
\end{align}

We follow the same chain of logic as in the zero temperature case and construct a solution as a perturbation in $A_0(r)$. Our idea is to find out a solution such that it approaches the scaling solution for $\frac{r}{r_0} \gg 1$. As we argue below, it would then be meaningful to match with the numerical solution to get a full solution of the EOMs.  Other quantities then automatically matches with the scaling solution.

The solution for $A_0$ in the above metric (eqn (\ref{testmetric})) is given by,
\begin{align}
A_0&= r_0^{2+\alpha} F(\frac{r}{r_0}) \\
F(r)&=\frac{\Gamma\left(\frac{2+\alpha}{3}\right) \Gamma\left(\frac{5+\alpha }{3}\right) }{\Gamma\left(\frac{2}{3}\right)\Gamma\left(\frac{5}{3}+\frac{2 \alpha }{3}\right)\sin(\frac{\pi \alpha}{3})}\frac{1}{r} \text{Im}\left(\text{\2F1}\left(-1-\frac{\alpha}{3},\frac{2}{3}+\frac{\alpha }{3},\frac{2}{3},r^3\right)\right)
\label{ftemp}
\end{align}
 Here, $A_0$ vanishes linearly near the black hole horizon.  For $r\rightarrow \infty$, $F(r)\approx r^{2+\alpha}+O(r^{1+\alpha})$. From that we get $A_0(r)\approx r^{2+\alpha}(1+r_0 O(\frac{1}{r}))$ for $r \gg r_0$. 
 
In a intermediate region $r=r_* \gg r_0$, the solutions in Eq. (\ref{ftemp}) could be matched with the zero temperature solution of $A_0$. Now if $r_0 \ll 1$, then we can choose the matching region such that $r_0 \ll r_* \ll 1$. Importantly $A_0$ remains small in the matching region and other fields could be solved perturbatively  in terms of $A_0$ (see appendix \ref{app:lim}). We get,
\begin{align}
\psi_1(r)&=-q^2 \psi_0 \int^r_{r_0}  \frac{d\tilde r}{g\tilde r^2} \int^{\tilde r}_{r_0}  \frac{r'^2 A_0^2}{g} dr' \\
\chi_1(r)&=-q^2 \psi_0^2 \int^r_{r_0}  \tilde r  \frac{A_0^2}{g^2} d\tilde r \\
g_1(r)&=- \frac{1}{r} \int^r_{r_0} \tilde r^2 \frac{A_0'^2}{4} d\tilde r.
\label{perturb}
\end{align}
Using the asymptotic expansion of the hypergeometric function one finds that the above quantities approach their zero temperature values in the matching region $r \sim r_*$. 

Now the solution at non-zero temperature could be integrated out to infinity. As our solution matches with the zero temperature solution in the leading order in $\frac{r_0}{r_*} \ll 1$, an almost same numerical solution may be used to extend our solution to all values of $r$.  We assume that the EOMs are numerically stable in the region $(r_*,\infty)$ in a sense that a small perturbative change in the initial condition gives rise to a small change at infinity. 

\subsection{Some results}

We do some simple calculation using our finite temperature solution. At $T\rightarrow 0$ the horizon behaves like a black hole horizon situated deep inside a IR $AdS^4$ and various $AdS^4$ results are applicable at the leading order in $r_0$. The temperature $T$ of the black hole is given by,
\begin{align}
4\pi T= [g'(g\exp(-\chi_0))']^\frac{1}{2}|_{r=r_0} \approx 3 \exp(-\chi_0/2) r_0, \quad \text{for } r_0 \ll 1.
\end{align}
Hence $T \propto r_0$ for small $r_0$. Total entropy($S$)\footnote{Condensate does not have any entropy. Hence the entropy of the whole solution is same as the entropy of the non-supercondcuting part.} and mass($M$) of the non-superconducting part (i.e. for the black hole) vary as, 
\begin{align}
S \propto r_0^2 \propto T^2
\end{align}
and,
\begin{align}
M \propto r_0 ^3 \propto T^3.
\end{align}

We may define two kind of specific heat for our system. One is at fixed chemical potential($C_\mu$) and other is at fixed total charge($C_\rho$) \cite{Peeters:2009sr}. Here we will calculate 
\begin{align}
C_\mu \sim T \frac{\partial S}{\partial T},\quad \mu \text{ fixed.}
\end{align}
Say a small change $\delta r_0$ in $r_0$ changes the system from $(T,\mu)$ to $(T+\delta T,\mu+\delta \mu)$. The resulting system is equivalent to a system $(\frac{T+\delta T}{\mu+\delta \mu} ,\mu)$. Here $\delta \mu \propto \delta r_0$ and $\delta T \propto \delta r_0$. Hence,

\begin{align}
C_{\mu} \propto r_0 ^2 \propto T^2
\end{align}

The charge density of the non-superconducting part behaves like,
\begin{align}
\rho \propto r_0^2 A' \propto r_0 ^{3+\alpha} \propto T^{3+\alpha}.
\end{align}

\subsection{Conductivity}
To obtain the conductivity in a background, one solves for a linearized perturbation of the vector
potential $A_x$ in the same geometry. Assuming $A_x=a(r) \exp(i\omega t)$ we get in our case,
\begin{align}
a''+\left(\frac{g'}{g}-\frac{\chi'}{2}\right) a'+ \left(e^{\chi}(\frac{\omega^2}{g^2}-\frac{A'^2}{g})-2q^2 \psi^2\right)a=0
\end{align}
We define a new variable,
\begin{align}
d \tilde r= \frac{e^{\frac{\chi}{2}}}{g} dr.
\end{align}
In terms of this new variable,
\begin{align}
-\frac{d^2}{d\tilde r^2}a+V(\tilde r)a=\omega^2 a
\label{Aeq}
\end{align}
where,
\begin{align}
V(\tilde r)=g [A_0'^2+2 q^2 \psi^2 \exp(-\chi)]
\end{align}
This potential vanishes near $r=0$. The superconducting nature of the system is argued from the existence of a supercurrent solution. If we set $\omega=0$ and integrate $A_x$ from the horizon (with a regularity condition at the horizon), we are expected to get a non-trivial $A_x$. Existence of such a solution implies a $\delta$ function for the real part of conductivity at $\omega=0$ \cite{Basu:2008st,Horowitz:2009ij,Herzog:2008he}.

The field $a$ has the following asymptotic behaviours near the horizon ($\tilde{r} \rightarrow -\infty$) and the boundary ($\tilde{r} \rightarrow 0$):
\begin{eqnarray}
 a(\tilde{r} \rightarrow 0) \sim a_0^b + a_1^b \tilde{r} \\
 a(\tilde{r} \rightarrow \infty) = a_0 e^{i \omega \tilde{r}},
\end{eqnarray}
Here, we have chosen the incoming boundary condition near the horizon.
Conductivity is defined as follows,
\begin{equation}
 \label{conductivity}
 \sigma = -\frac{ia_1^b}{\omega a_0^b}
\end{equation}

It has been argued that at zero temperature, the zero frequency limit (i.e. the non-superconducting part) of ${\text Re}(\sigma)$ vanishes as powerlaw, i.e. \cite{Horowitz:2009ij,Gubser:2008pf},
\bea
{\text Re}(\sigma(\omega)) \sim \omega^\delta, \text{  for } \delta\ll 1.
\eea
Here $\delta=\sqrt{4 V_0+1}-1$.  Where $V_0=\lim_{\tilde r \rightarrow \infty} \tilde r^2 V(\tilde r)$. As the non-superconducting contribution to ${\text Re}(\sigma)$ is non-zero even at small frequencies, the system does not have a energy gap in this channel. However non-superconducting part of ${\text Re}(\sigma(\omega))$ vanishes at the zero frequency limit.

\subsubsection{Small non-zero $T$}
Non-superconducting part of $\lim_{\omega \rightarrow 0}{\text Re}(\sigma(\omega))$ is non-zero at any finite temperature. Due to the gaplessness of the system we expect a powerlaw decay of the above quantity with the temperature. The non-superconducting contribution to $\lim_{\omega \rightarrow 0}{\text Re}(\sigma(\omega))$ has a smooth zero frequency limit (see appendix \ref{app:cond}) and could be calculated by setting $\omega=0$ in \eref{Aeq}, i.e.
\begin{align} 
\frac{d^2}{d\tilde r^2} a=g \left ( A'(x)^2+q^2 e^{-\chi}A_0^2 \right)a  \\
\end{align}
and we have,
\begin{align} 
\lim_{\omega \rightarrow 0}{\text Re}(\sigma(\omega))=\frac{a_{h}^2}{a_b^2}.
\end{align}
Where,  $a_h$ and $a_b$ is the value of $a$ at the horizon and boundary respectively. We  break down the domain of 
$r$ in two parts $(r_0,r_*)$ and $(r_*,\infty)$, s.t., $r_0 \ll r_* \ll 1$. We have,

\begin{align} 
\lim_{\omega \rightarrow 0}{\text Re}(\sigma(\omega))=\frac{a_{h}^2}{a_b^2}=\frac{a_{h}^2}{a(r_*)^2}\frac{a(r_*)^2}{a_b^2}.
\end{align}

In terms of the new co-ordinate $r_{*}=\tilde r$. Our goal is to fix $r_*$ and take $r_0$ to a zero. In that case, 
\begin{align} 
\lim_{\omega \rightarrow 0}{\text Re}(\sigma(\omega))=\frac{a_{h}^2}{a_b^2} \sim \frac{a_{h}^2}{a(r_*)^2} C_1
\end{align}
Where $C_1$ is the limiting value of the quantity $\frac{a(r_*)^2}{a_b^2}$ as $r_0 \rightarrow 0$. This value may be calculated from the numerics. Leading dependence of non-superconducting part of $\mathrm{Re}(\sigma(0))$ on $\frac{1}{T}$ comes from the behaviour of the solution between $(r_0,r_*)$. Taking $r_*$ in our matching region we can use our analytic solution in the matching region. Defining a rescaled variable $r_1=\frac{r}{r_0}$ and the corresponding rescaled variable $\tilde r_1=\tilde r r_0$, we write the equations \eref{Aeq} as,
\begin{align}  
\frac{d^2}{d\tilde r_1^2} a&= r_1^2 (1-\frac{1}{r_1^3}) \left ( c^2 A'(x)^2+2 q^2 e^{-\chi}\psi^2 \right)a \\
&\approx 2q^2 r_1^2 (1-\frac{1}{r_1^3}) e^{-\chi_0}\psi_0^2  a
\end{align}
where we have kept the leading order terms in $r_0$.
The regular solution at the horizon has the following form,
\begin{align}
a=\text{Im}\left[\text{\2F1}\left(-\frac{\alpha }{3}-\frac{2}{3},\frac{\alpha
   }{3}+1,\frac{1}{3},\frac{r^3}{r_0^3}\right)\right].
\end{align}

Using the asymptotic expansion of the hypergeometric function we get,
\bea
\lim_{\omega \rightarrow 0}{\text Re}(\sigma(\omega)) \sim T^{2+\alpha}
\eea
\section{Non-abelian case}
\label{sec:nonab}

The Einstein-YM action for a non-abelian gauge field with a negative cosmological constant is given by \cite{Gubser:2008zu},
\begin{eqnarray}
{\cal L} =\int d^4x\sqrt{-g}\left({\mathcal
{R}}+\frac{6}{l^2}-\frac{1}{4}F_a^{\mu\nu}F_{\mu\nu}^a\right),
\label{actionem}
\end{eqnarray}
where $F_{\mu\nu}$ is the field strength of an $SU(2)$ gauge field. The fully backreacted solution of the above equations is constructed in \cite{Ammon:2009xh,Basu:2009vv}. 

The ansatz for the gauge fields is\footnote{Due to a repulsive term coming from the non-abelian interactions, it is expected that a isotropic ansatz will have a quartic instability and would possibly have more free energy than the anisotropic ones \cite{Gubser:2008wv,Basu:2008bh}.},
\begin{eqnarray}
A=A(r)\tau^3 dt+B(r)\tau^1 dx.
\end{eqnarray}

To tally with the anisotropy of the gauge field ansatz in the spatial direction, we choose the following ansatz for our metric,
\begin{eqnarray}
ds^2=-g(r)e^{-\chi(r)}dt^2+\frac{dr^2}{g(r)}+r^2\Big(c(r)^2 dx^2+dy^2\Big).
\end{eqnarray}


The Maxwell's equations of $A(r),B(r)$ are
\begin{eqnarray}
A_t^3\longrightarrow& A''+A'\left(\frac{2}{r}+\frac{\chi'}{2}+\frac{c'}{c}\right)-\frac{q^2B^2}{r^2gc}A=0, \nonumber\\
A_x^1\longrightarrow& B''+B'\left(\frac{g'}{g}-\frac{\chi'}{2}-\frac{c'}{c}\right)+\frac{e^{\chi}q^2A^2}{g^2}B=0.
\label{maineq}
\end{eqnarray}

 The diagonal Einstein equations give,
\begin{eqnarray}
  -g'\left(\frac{1}{r}+\frac{c'}{2c}\right)-g\left(\frac{1}{r^2}+\frac{3c'}{r c}+\frac{c''}{c}\right)+3
   &=& \frac{e^{\chi}}{4}A'^2+\frac{g}{4r^2c}B'^2+e^{\chi}\frac{q^2A^2B^2}{4r^2gc},  \nonumber\\
  -\frac{\chi'}{r}+\frac{c'}{c}\left(-\chi'+\frac{g'}{g}\right) &=& \frac{e^{\chi}q^2A^2B^2}{g^2r^2c^2}, \nonumber\\
  cc''+cc'\left(\frac{g'}{g}+\left(\frac{2}{r}-\frac{\chi'}{2}\right)\right) &=& -\frac{B'^2}{2r^2}+e^{\chi}\frac{q^2A^2B^2}{2g^2r^2}.
\label{maineq2}
\end{eqnarray}

The above equations are invariant under the following scaling symmetries:
\begin{eqnarray}
\label{rescale}
& &r \rightarrow a_1 r, \quad (t,x,y) \rightarrow (t,x,y)/a_1, \quad g \rightarrow a_1^2g, \quad A \rightarrow a_1 A, \quad B \rightarrow a_1 B, \\
\nn & &e^\chi \rightarrow a_2^2 e^\chi, \quad t \rightarrow a_2 t, \quad A \rightarrow A/a_2. \\
\nn & & x \rightarrow x/a_3,\quad B \rightarrow a_3B. \quad c \rightarrow a_3 c.
\end{eqnarray}
The second scaling symmetry may be used to set $\chi=0$ at infinity and the third scaling symmetry may be used to set $c=1$ at infinity, so that the asymptotic metric is that of $AdS_4$.

The fields have the following asymptotic behavior:
\begin{equation}
A = \mu - \frac{\rho}{r}, \quad B = B_0^b + \frac{B_1^b}{r},
\label{asymptotic}
\end{equation}
where $\mu$ is the chemical potential and $\rho$ is the charge density in the boundary theory. In what follows we will only consider the solutions for the field $B$ which vanishes near the boundary, i.e. $B_0 = 0$.

\subsection{Zero temperature solution}
\label{sec:nonabzero}
Like the abelian case the zero temperature solution is constructed by similar techniques \cite{Basu:2009vv}. We start by guessing near horizon ansatz $g=r^2,B= B_0 $,
 Putting this in \eref{maineq}, we get find out the equation of motion for $A$,
\bea
r^2 (r^2 A')'=\frac{q^2 B_0^2} {A}
\Rightarrow A = e^{-\frac{\beta}{r}}, \quad \beta=q B_0/{c_0},
\eea
where we have used the observation $A \rightarrow 0$ at the horizon and by rescaling (\ref{rescale}) we set the coefficients $A_0=1$. In the next order in the perturbation $A_0$, we get, 
\begin{eqnarray}
~~B\sim B_0-B_1(r) ,~~\chi\sim \chi_0-\chi_1(r),~~g\sim r^2+g_1(r),~~c\sim c_0+c_1(r) .
\label{ansatz}
\end{eqnarray}
All the terms with subscript $1$ are sub-leading and go to zero, faster than the leading part where it is applicable, as $r\rightarrow0$. Here,
\begin{eqnarray}
 ~~~B_1= B_0\left(\frac{e^{\chi_0}q^2 }{4\beta^2}e^{-2\beta/r}\right),
~~~c =c_0\left(\frac{e^{\chi_0} }{8r^2}e^{-2\beta/r}\right), \\
\nn \chi_1=-\frac{e^{\chi_0}}{2r}e^{-2\beta/r},~~~g_1= -\frac{e^{\chi_0}A_0^2\beta}{4r}e^{-2\beta/r},
\label{scaling2}
\end{eqnarray}
where by rescaling (\ref{rescale}) we may set $\chi_0=0,c_0=1$. After one solves this equation by numerics, one again uses the rescalings of $g,c,\chi$ to make the Asymptotic geometry the same as that of $AdS_4$. For a given $q$, one numerically choose $\beta$ in such a fashion that $B$ vanishes near the boundary \cite{Basu:2009vv}.

\subsection{Small non-zero $T$}

We follow the same strategy as in the abelian case and choose a finite temperature metric and $B$ field like \footnote{In a similar discussion of a non-zero temperature solution, the metric used was not correct in \cite{Basu:2009vv} and other formulas are also schematic.},

\bea
g(r)=r^2 \left(1-\frac{r_0^3}{r^3}\right), \quad B=B_0
\eea
This satisfies the Einstein's equations at the zero'th order. Using this background fields we may write down the equation for $A$,

\bea
A''+\frac{2}{ r}A'-\frac{\beta^2}{  r^4 (1-r_0^3/r^3)}A=0,\quad \beta=\frac{q B_0}{c_0}
\label{nonabA}
\eea
We would like to find a solution to the above equation which is regular at the horizon and approaches the zero temperature solution $\exp(-\frac{\beta}{r})$ for $r \gg r_0$. Unfortunately there seems to be no analytic solutions to the above equation. However, any solution $A(r)$ may be written as a linear combination
\begin{align}
A(r)=C_1 \exp(-\frac{\beta}{r}) + C_2 \exp(\frac{\beta}{r}) \text{ for } r \gg r_0.
\end{align}
We need to show that as $r_0 \rightarrow 0$, we may choose $r_* (\text{where } r_* \gg r_0)$ in such a way that $C_2/C_1  \ll \exp(2\frac{\beta}{r_*})$. Just like the abelian case we may choose a $r_*$, s.t. $r_0 \ll r_* \ll 1$. So that the nonzero-temperature solution approaches the zero temperature solution in the regime $r \sim r_*$. This amounts to saying that \eref{nonabA} has a smooth zero temperature limit, at least for the solutions which are regular at the horizon. This is argued using matched asymptotic expansion. Near the horizon $r\approx r_0+\delta r, \delta r \ll 1$ and it is possible to linearise \eref{nonabA}. The  solution of the linearised equation which is regular at the horizon is given by,
\begin{align}
A(r)\approx \frac{r_0^{3/2}}{\sqrt{\delta r}} {\text I}_1\left(\beta \frac{2 \sqrt{\delta r}}{\sqrt{3} {r_0}^{3/2}}\right)
\end{align}
For small enough $r_0$ there is a region where both the above linear approximation and WKB solution of \eref{nonabA} are both valid. Moreover one may choose $ \frac{r_0^{3/2}}{\sqrt{\delta r}} \gg 1$ in such a region. Using the asymptotic expansion of the Bessel function one argues that the regular solution matches with the correct WKB solution. Extrapolating the correct solution to $r \gg r_0$ one gets,
\begin{align}
A(r)\approx C_1 \exp(-\frac{\beta}{r})  \text{ for } r \gg r_0.
\end{align}
This guarantees a matching region where the non-zero temperature solution approaches the zero temperature solution. Other fields may be solved following the similar procedure to that of abelian case. Using the similar argument of numerical stability we expect that our solution may be integrated out to infinity using a slight variation of the zero temperature numerics.

\subsection{Some results}
We will make some simple calculation using our scaling solution. As the black hole becomes a small black hole situated deep inside a IR $AdS^4$ various $AdS^4$ results are applicable at the leading order in $r_0$. The temperature $T$, entropy, mass and specific heat follow the similar behaviour to that of abelian case.
The charge density of the non-superconducting part behaves like,
\begin{align}
\rho \propto r_0^2 A' \propto \exp(-\frac{\beta}{r_0}) \sim \exp(-\frac{\beta}{T}) .
\end{align}

\subsection{Conductivity and energy gap}

In order to calculate the conductivity of this system, we need to turn on a small perturbation in the vector potential. We turn on the gauge field perturbations of the form:
\begin{equation}
A_y^3 = \epsilon a(r)e^{-i\omega t}\tau^3 dy
\end{equation}
Here we get,
\begin{eqnarray}
\label{nonabcond}
a''+a'\left(\frac{g'}{g}-\frac{\chi'}{2}+\frac{c'}{c}\right)
  +a\left(\frac{e^{\chi}\omega^2}{g^2}-\frac{q^2B^2}{gr^2c^2} -e^{\chi}\frac{A'^2}{g}\right)=0.
\end{eqnarray}

This can be written as a Schr\"{o}dinger equation:
\begin{equation}
 \label{schroedinger}
 -a'' + V(\tilde{r})a = c^2 \omega^2 a,
\end{equation}
where
\begin{eqnarray}
V(r)=g\left(c^2A'^2+e^{-\chi}\frac{q^2B^2}{r^2}\right).
\label{vr}
\end{eqnarray}
and all the derivatives in \eref{schroedinger} are in terms of the new variable new variable $\tilde{r}$ (``tortoise coordinate'') given by:
\begin{equation}
 \frac{d}{d\tilde{r}} \equiv e^{-\chi/2} gc \frac{d}{dr}.
\label{tortoise}
\end{equation}
In terms of the new co-ordinate horizon is mapped to $\tilde r=-\infty.$

Here, $c$  approaches to unity as $r \rightarrow \infty$, so that the spacetime is asymptotically $AdS_4$. It follows then from  \eref{asymptotic} that the potential $V(r)$ vanishes near the boundary. If we require $g\sim r^2$ near the horizon at $r=0$ then the first term vanishes, while the second term is finite as $B(r=0) \equiv B_0 \neq 0$ and $\chi$ is also finite at the horizon. Note that since $c \rightarrow 1$ near the boundary, the quantity $\omega$ can be interpreted as the frequency of the incoming wave.

Similar to abelian case, the superconducting nature of the system is argued from the existence of a supercurrent solution. 

The nature of the finite part of the conductivity can be inferred from the potential $V(r)$ and shows a hard gap \cite{Basu:2009vv} at $T=0$. The fact that the potential is nonzero at the horizon at $T=0$ makes it possible for this system to exhibit a hard gap. From \eref{schroedinger}, the field $a$ has the following asymptotic behaviours near the horizon ($\tilde{r} \rightarrow -\infty$) and the boundary ($\tilde{r} \rightarrow 0$):
\begin{eqnarray}
 a(\tilde{r} \rightarrow 0) \sim a_0^b + a_1^b \tilde{r} \\
 a(\tilde{r} \rightarrow \infty) = a_0 e^{i \tilde{\omega} \tilde{r}},
\end{eqnarray}
where $\tilde{\omega} = \sqrt{c_0^2 \omega^2 - V_0}$, with $c_h$, $a_h$ being the near-horizon values of $c$ and $a$ respectively. Here, we have chosen the incoming boundary condition near the horizon. The conductivity of the system is defined in the same way as in \eref{conductivity}.

It follows from \eref{schroedinger} that :
\begin{equation}
 a^* a'' - a a^{*''} = 0,
\end{equation}
which implies that the quantity $\Lambda = a^* a' - a a^{*'} = 2i \mathrm{Im}(a a^{*'})$ is a constant.

\subsubsection{$T=0$ case}
 At T=0 , equating the values of $\Lambda$ near the horizon and the boundary we get:
\begin{equation}
 \mathrm{Re}(\sigma) = \left\{
\begin{array}{l l}
\frac{\tilde{\omega}}{\omega} \frac{|a_0|^2}{|a_0^b|^2} & , \quad \tilde \omega^2 > 0 \\
                         0,& \quad  \tilde \omega^2 < 0 \\
\end{array} \right.
\label{finalcond}
\end{equation}
Therefore, the real part of the conductivity will vanish whenever $\tilde{\omega}$ is imaginary, i.e. when $\omega < \Delta_1=\sqrt{V_0}/c_0$, which defines the gap.

\subsubsection{Small nonzero $T$}
However at any finite temperature $T$, the system does not show a hard gap as the potential $V(r)$ actually vanishes near the black hole horizon. This is an expected behaviour considering thermal excitations of the condensate. One may define another gap by the low temperature behaviour of the conductivity. Considering the finite part (i.e. non-superconducting contribution) of  $\mathrm{Re}(\sigma)$ at zero frequency limit at low temperature, one expects
\begin{align}
\lim_{\omega\rightarrow 0} \mathrm{Re}(\sigma(\omega)) \sim \exp(-\frac{\Delta_2}{T}).
\end{align}
Generically $\Delta_2 \neq \Delta_1$ and their ratio gives information about the pairing mechanism. $\Delta_2$ may be thought as the mass of the charged quassiparticle carriers in the system. $\Delta_1$ may be thought as the mass of the "pairs//combination" of the quassiparticles which gives excitation over the pure condensate. In BCS theory such a combination of basic carriers is a 'Cooper pair'. In the BCS theory, $\Delta_1=2 \Delta_2$. We would like to calculate $\Delta_2$ from our low temperature solution.

The non-superconducting contribution to $\mathrm{Re}(\sigma)(0)$ has a smooth zero frequency limit and could be calculated by setting $\omega=0$ in \eref{schroedinger}, i.e.
\begin{align} 
\label{sch2}
\frac{d^2}{d\tilde r^2} a=g \left ( c^2 A'(x)^2+q^2 e^{-\chi}\frac{B^2}{r^2}\right)a  \\
\end{align}
and we have,
\begin{align} 
\lim_{\omega\rightarrow 0} \mathrm{Re}(\sigma(\omega))=\frac{a_{h}^2}{a_b^2}.
\end{align}
Where,  $a_h$ and $a_b$ is the value of $a$ at the horizon and boundary respectively. We  break down the domain of 
$r$ in two parts $(r_0,r_*)$ and $(r_*,\infty)$, s.t., $r_0 \ll r_* \ll 1$. We have,

\begin{align} 
\lim_{\omega\rightarrow 0} \mathrm{Re}(\sigma(\omega)) =\frac{a_{h}^2}{a_b^2}=\frac{a_{h}^2}{a(r_*)^2}\frac{a(r_*)^2}{a_b^2}.
\end{align}

In terms of the new co-ordinate $r_{*}=\tilde r$. Our goal is to fix $r_*$ and take $r_0$ to a zero. In that case, 
\begin{align} 
\lim_{\omega\rightarrow 0} \mathrm{Re}(\sigma(\omega))=\frac{a_{h}^2}{a_b^2} \sim \frac{a_{h}^2}{a(r_*)^2} C_1
\end{align}
Where $C_1$ is the limiting value of the quantity $\frac{a(r_*)^2}{a_b^2}$ as $r_0 \rightarrow 0$. This value may be calculated from the numerics. Leading dependence of non-superconducting part of $\mathrm{Re}(\sigma(0))$ on $\frac{1}{T}$ comes from the behaviour of the solution between $(r_0,r_*)$. Taking $r_*$ in our matching region we can use our analytic solution in the matching region. Defining a rescaled variable $r_1=\frac{r}{r_0}$ and the corresponding rescaled variable $\tilde r_1=\tilde r r_0$, we write the equations \eref{sch2} as,
\begin{align}  
\frac{d^2}{d\tilde r_1^2} a&=\frac{1}{r_0^2} r_1^2 (1-\frac{1}{r_1^3}) \left ( c^2 A'(x)^2+q^2 e^{-\chi}\frac{B^2}{r_1^2}\right)a \\
&\approx q^2 \frac{1}{r_0^2} r_1^2 (1-\frac{1}{r_1^3}) e^{-\chi_0}\frac{B_0^2}{r_1^2} a
\end{align}
where we have kept the leading order terms in $r_0$. 
For $r_0\ll 1$ the above equation may be solved using WKB approximation. One may question of validity of WKB approximation as the potential vanishes near the black hole horizon. This turns out not be a problem as we can again break down the range of $r_1$ of into two parts $[1,r_2]$ and $[r_2,r_*)$. We choose our $r_2$, s.t. at $r\sim r_2$ WKB solution is valid. For a small $r_0$, $r_2$ lies very close to horizon $r_1=1$. For example one may choose $r_2=1+\sqrt{r_0}$. We can also use near horizon linearisation of metric and solve $a(r_1)$ terms of Bessel functions for $r_1<r_2$,
\bea
a(r_1)={\rm I}_0\left(\frac{q B_0 e^{-\frac{\chi_0}{2}}\sqrt{r_1-1}}{r_0}\right)
\eea

 At $r \sim r_2$ both the Bessel function and WKB method is valid and we can match these two following the principle of matched asymptotic expansion.  Using this method one finds out that in the leading order only the WKB contribution matters,
\begin{align} 
\lim_{\omega\rightarrow 0} \mathrm{Re}(\sigma(\omega)) &\sim  \exp\left(-2 \frac{q B_0}{c_0 r_0} \int_{1}^{\infty} \frac{dr}{r^2 \sqrt{1-\frac{1}{r^3}}}\right) 
\end{align}
In the above we also take $\frac{r_*}{r_0}\rightarrow \infty$ and $\frac{r_2}{r_1}\rightarrow 1$ limit.

Using the formula for temperature we get,
\begin{align} 
\Delta_2= \Delta_1 \frac{3}{2\pi}\int_{1}^{\infty} \frac{dr}{r^2 \sqrt{1-\frac{1}{r^3}}} = \Delta_1  \frac{3\Gamma\left(\frac{4}{3}\right)}{2\sqrt{\pi}\Gamma\left(\frac{5}{6}\right)} \approx 0.669 \Delta_1 
\end{align}
This may be contrasted with the weak coupling BCS value of $\Delta_2= \frac{1}{2}\Delta_1$. We find around $33\%$ deviation from the BCS value. Interestingly the numerical factor is close to $\frac{2}{3}$. This gives us information about possible underlying strong coupling pairing mechanism \cite{Hartnoll:2008vx}.


\section{Acknowledgements}
I thank Jianyang He, Moshe Rozali and Sumit Das for various discussions. I thank people of University of BC and University of KY for support. I am supported by grant NSF-PHY-0855614 and NSF-PHY-0970069. 
\appendix
\section{Conductivity at $\omega=0$}
\label{app:cond}
Let's consider a equation similar to \eref{Aeq},
\begin{align}
-\frac{d^2}{d r^2}a+V(r)a=\omega^2 a
\label{Aeq2}
\end{align}

 with $r\in[0,\infty]$ and the potential vanishes near the horizon ($r=\infty$)\footnote{Here $r$ is same as co-ordinate $\tilde r$ in the main text.  $\tilde r$ is not used in the appendix due to notational simplicity.} The claim is that,

\bea
\lim_{\omega \rightarrow 0} \text{Re} \sigma(\omega)= \frac{|a(\infty)^2|}{|a(0)^2|}
\label{toprove}
\eea 
where we find $a(r)$ by solving  \eref{Aeq2} with $\omega=0$ and using the regularity boundary condition at the horizon. 
 
From \eref{Aeq2}, field $a$ has the  following behaviour near boundary ($r \rightarrow 0$):
\bea
 a(\tilde{r} \rightarrow 0) \sim a(0) + a'(0) \tilde{r} \\
\eea

The conductivity of the system is defined as,
\bea
 \sigma(\omega)= \frac{a'(0)}{i\omega a(0)}
\eea
Here, we have chosen the incoming boundary condition near the horizon.

It follows from \eref{Aeq2} that :
\bea
 a^* a'' - a a^{*''} = 0,  , 
\eea
which implies that the quantity $\Lambda = a^* a' - a a^{*'} = 2i \mathrm{Im}(a a^{*'})$ is a constant.

Equating the values of $\Lambda$ near the boundary and any intermediate distance $r1$ we get:
\begin{equation}
 \mathrm{Re}(\sigma(\omega)) = \frac{\mathrm{Im}(a^{*}(r1)a'(r1))}{\omega |a(0)|^2}
\end{equation}

Now, for small enough $\omega$ one may break down the range of $r$ into two parts $[0,r_*],[r_*,\infty]$. Where $r_*$ is such that $V(r) \gg \omega^2 $ for $r < r_*$ and $\omega$ could be treated as a small perturbation in this regime. Also, $r_* \gg 1$ such that only leading terms contributes in $V(r)$ for $r > r_*$. In the second region we use,
\bea
V(r) \approx C_1 \exp(-C_2 r)
\eea
We find,
\bea
a(r)=\text{I}_{\frac{2 i \omega}{C_2}}\left(\frac{2 \sqrt{C_1} \sqrt{e^{-r C_2 }}}{C_2}\right).
\eea
Where we have chosen the solution with correct incoming condition near $r=\infty$. This solution approaches the regular solution for $\omega=0$, $\text{I}_{0}\left(\frac{2 \sqrt{C_1} \sqrt{e^{-r C_2 }}}{C_2}\right)$ near $r\sim r_*$. Also as $r_* \gg 1$, we may use the asymptotic form of the Bessel function to get,
\bea
a(r) \sim e^{i r \omega},\quad r \sim r_*.
\eea
Hence conductivity is given by,
\begin{equation}
 \mathrm{Re}(\sigma(\omega)) = \frac{|a(r_*)^2|}{|a(0)^2|}
\end{equation}
Our assertion \eref{toprove} is proved by noticing that as $\omega \rightarrow 0$ ,  $r_* \rightarrow \infty$.  

\section{Perturbation in $A_0$}
\label{app:lim}
At a more technical level one may ask why we neglect the back reaction of \eref{perturb} on the EOM's, i.e. what exactly we mean by $A_0$ small. For example to neglect the back reaction of $\chi_1'$ one must have $\chi_1' r \ll 1$.,
\begin{align}
 \chi_1' r
&= r^2 r_0^{4+2\alpha}\frac{ F(\frac{r}{r_0})^2}{r^4(1-\frac{r_0}{r^3})^2} \\
&=  r^{2+2\alpha} a^{2-2\alpha} \frac{F(a)^2}{(a^3-1)^2}, \\
&= r^{2+2\alpha} C_1 \ll 1, \text{ where } C_1 \text{ is a constant.}
\end{align}
Where we define $a=\frac{r}{r_0}$. In our limit $r \ll 1$ although $a$ may be large. In the last line we have used the the property that $a^{2-2\alpha} \frac{F(a)^2}{(a^3-1)^2}$ is bounded function in the domain $[1,\infty]$, i.e. there exists $C_1>0$ such that $a^{2-2\alpha} \frac{F(a)^2}{(a^3-1)^2}<C_1$ for $a\in[1,\infty]$.
Using similar techniques one also justify the perturbative expansion for other quantities.

\bibliographystyle{JHEP}
\bibliography{nonab}
\end{document}